\documentclass[conference]{IEEEtran}
\IEEEoverridecommandlockouts
\usepackage{cite}
\usepackage{amsmath,amssymb,amsfonts}
\usepackage{algorithmic}
\usepackage{graphicx}
\usepackage{textcomp}
\usepackage{xcolor}
\usepackage{hyperref}
\usepackage{balance}
\usepackage{algorithm}
\usepackage{algorithmic}
\usepackage{amsmath}
\usepackage{booktabs}
\usepackage{amsfonts}
\usepackage{microtype}  
\usepackage[list=true]{subcaption}
\usepackage[font={footnotesize}]{caption}
\usepackage{times}
\usepackage{graphicx}
\usepackage{multirow}
\usepackage[none]{hyphenat}
\usepackage{float}
\usepackage{acro}
\usepackage{tikz}
\usepackage{pgfplots}
\usetikzlibrary{svg.path}
\tikzset{every picture/.style={line width=0.6pt}}
\usetikzlibrary{calc,positioning}
\usetikzlibrary{arrows.meta,
                backgrounds,
                chains,
                fit,
                quotes}
\usepackage[bottom]{footmisc}

\newtheorem{proposition}{Proposition}
                
\DeclareAcronym{5g}{
short=5G,
long= fifth generation,
}
\DeclareAcronym{iq}{
short=I/Q,
long= quadrature,
}
\DeclareAcronym{I}{
short=I,
long= in-phase,
}
\DeclareAcronym{Q}{
short=Q,
long= quadrature,
}
\DeclareAcronym{ls}{
short=LS,
long= least squares,
}
\DeclareAcronym{ota}{
short=OTA,
long= over-the-air,
}
\DeclareAcronym{sgd}{
short=SGD,
long= stochastic gradient descent,
}
\DeclareAcronym{ce}{
short=CE,
long= cross-entropy,
}
\DeclareAcronym{sl}{
short=SL,
long= supervised learning,
}
\DeclareAcronym{rl}{
short=RL,
long= reinforcement learning,
}
\DeclareAcronym{awgn}{
short=AWGN,
long= additive white Gaussian noise,
}
\DeclareAcronym{ser}{
short=SER,
long= symbol error rate,
}
\DeclareAcronym{qam}{
short=QAM,
long= quadrature amplitude modulation,
}
\DeclareAcronym{rrc}{
short=RRC,
long= root-raised cosine,
}

\DeclareAcronym{snr}{
short=SNR,
long= signal-to-noise ratio,
}
\DeclareAcronym{rvftdnn}{
short=RVFTDNN,
long= real-valued focused time-delay neural network,
}
\DeclareAcronym{lo}{
short=LO,
long= local oscillator,
}
\DeclareAcronym{lpf}{
short=LPF,
long= lowpass filter,
}
\DeclareAcronym{pdf}{
short=PDF,
long= probability density function,
}
\DeclareAcronym{cdf}{
short=CDF,
long= cumulative distribution function ,
}
\DeclareAcronym{fir}{
short=FIR,
long= finite impulse response,
}
\DeclareAcronym{rhs}{
short=RHS,
long= right-hand side,
}
\DeclareAcronym{dsp}{
short=DSP,
long= digital signal processing,
}
\DeclareAcronym{nn}{
short=NN,
long= neural network,
}
\DeclareAcronym{mlp}{
short=MLP,
long=multilayer perceptron
}
\DeclareAcronym{GaN}{
short=GaN,
long=Gallium Nitride,
}
\DeclareAcronym{relu}{
short=ReLU,
long = rectified linear unit, 
}
\DeclareAcronym{mse}{
short=MSE,
long=mean squared error,
}
\DeclareAcronym{rvtdnn}{
short=RVTDNN,
long= real-valued time-delay neural network,
}
\DeclareAcronym{arvtdnn}{
short=ARVTDNN,
long= augmented real-valued time-delay neural network,
}
\DeclareAcronym{arden}{
short=ARDEN,
long= attention residual real-valued time-delay neural network,
}
\DeclareAcronym{r2tdnn}{
short=R2TDNN,
long= residual real-valued time-delay neural network,
}
\DeclareAcronym{flop}{
short=FLOP,
long= floating point operations,
}
\DeclareAcronym{ph}{
short=PH,
long= parallel Hammerstein
}
\DeclareAcronym{ofdm}{
short=OFDM,
long=orthogonal frequency division multiplexing,
}
\DeclareAcronym{par}{
short=PAR,
long=peak-to-average ratio,
}
\DeclareAcronym{papr}{
short=PAPR,
long=peak-to-average power ratio,
}
\DeclareAcronym{rf}{
short=RF,
long=radio frequency,
}
\DeclareAcronym{pa}{
short=PA,
long=power amplifier,
}
\DeclareAcronym{pas}{
short=\acs{pa}s,
long=power amplifiers,
}
\DeclareAcronym{psd}{
short=PSD,
long= power spectral density,
}
\DeclareAcronym{dpd}{
short=DPD,
long=digital predistortion,
}
\DeclareAcronym{cfr}{
short=CFR,
long=crest factor reduction,
}
\DeclareAcronym{cf}{
short=CF,
long=crest-factor}

\DeclareAcronym{evm}{
short=EVM,
long=error vector magnitude,
}
\DeclareAcronym{nmse}{
short=NMSE,
long=normalized mean squared error,
}
\DeclareAcronym{acpr}{
short=ACPR,
long=adjacent channel power ratio,
}
\DeclareAcronym{pae}{
short=PAE,
long=power added efficiency,
}
\DeclareAcronym{dla}{
short=DLA,
long=direct learning architecture,
}
\DeclareAcronym{ila}{
short=ILA,
long=indirect learning architecture,
}
\DeclareAcronym{ilc}{
short=ILC,
long=iterative learning control ,
}
\DeclareAcronym{cfr-dpd}{
short=CFR-DPD,
long=CFR combined with DPD,
}
\DeclareAcronym{icf}{
short=ICF,
long=iterative clipping and filtering,
}
\DeclareAcronym{am/am}{
short=AM/AM,
long=amplitude-to-amplitude,
}
\DeclareAcronym{am/pm}{
short=AM/PM,
long=amplitude-to-phase,
}
\DeclareAcronym{siso}{
short=SISO,
long=single-input single-output
}
\DeclareAcronym{mimo}{
short=MIMO,
long=multiple-input multiple-output
}
\DeclareAcronym{mp}{
short=MP,
long=memory polynomial
}
\DeclareAcronym{gmp}{
short=GMP,
long=generalized memory polynomial
}
\DeclareAcronym{adc}{
short=ADC,
long= analog-to-digital converter}
\DeclareAcronym{dac}{
short=DAC,
long= digital-to-analog converter}
\DeclareAcronym{ilc-dpd}{
short=ILC-DPD,
long= adaptive ILC-based DPD
}
\DeclareAcronym{rms}{
short=RMS,
long= root mean squares
}
\DeclareAcronym{vst}{
short=VST,
long= vector signal transceiver
}
\DeclareAcronym{mmwv}{
short=mm-Wave,
long= millimeter-wave
}

\begin{document}
\bstctlcite{IEEEexample:BSTcontrol}

\title{Symbol-Based Over-the-Air Digital Predistortion  \\ Using Reinforcement Learning
\\
\thanks{This work was supported by the Swedish Foundation for Strategic Research (SSF), grant no.~I19-0021.}}

\author{Yibo~Wu\IEEEauthorrefmark{1}\IEEEauthorrefmark{2},
Jinxiang~Song\IEEEauthorrefmark{2}, Christian~H{\"a}ger\IEEEauthorrefmark{2},
        Ulf~Gustavsson\IEEEauthorrefmark{1},\\
        Alexandre~Graell~i~Amat\IEEEauthorrefmark{2}, and
        Henk~Wymeersch\IEEEauthorrefmark{2}\\
        \IEEEauthorrefmark{1}Ericsson Research, Gothenburg, Sweden\\
        \IEEEauthorrefmark{2}Department of Electrical Engineeering, Chalmers University of Technology, Gothenburg, Sweden
        }

\maketitle

\begin{abstract}
We propose an \acl{ota} \acl{dpd} optimization algorithm using \acl{rl}. Based on a symbol-based
criterion, the algorithm minimizes the errors between downsampled messages at the receiver side. The algorithm does not require any knowledge about the underlying hardware or channel. For a generalized memory polynomial power amplifier and additive white Gaussian noise channel, we show that the proposed algorithm achieves performance improvements in terms of symbol error rate compared with an indirect learning architecture even when the latter is coupled with a full sampling rate ADC in the feedback path. Furthermore, it maintains a satisfactory \acl{acpr}.
\end{abstract}

%
\section{Introduction}
\Acf{dpd} is a technique to linearize the nonlinear \acf{pa} in a \acf{rf} chain to achieve the best energy efficiency while maintaining the spectral mask requirement~\cite{cripps2006rf}. It is customary to implement \ac{dpd} using parametric models. For simplicity, \ac{dpd} parameters are mostly optimized at the transmitter side, which requires a feedback data acquisition path to collect the PA output signal~\cite{liu2018beam}. To capture the full-band behavior of the PA for DPD optimization, high sampling rate feedback \acp{adc} in the feedback path are needed, which is challenging for wideband signals. High sampling rate \acp{adc} add a huge cost, which increases linearly with the number of \ac{rf} chains in massive \ac{mimo}~\cite{larsson2014massive}, where each RF chain may require a separate feedback path~\cite{liu2018beam}.

To tackle the high cost problem of the feedback path, recent works have shifted toward low sampling rate ADC methods~\cite{liu2014novel,
wang2016low,guan2017digital,beltagy2018direct}, where the feedback path is coupled with a low sampling rate ADC. These works focus on recovering the full-rate PA output signal from the undersampled output signal of the low-rate ADC, and the recovered signal is then used for DPD optimization. To reduce the cost of the feedback path further, \acf{ota} methods have become a promising solution for the DPD optimization~\cite{hausmair2018modeling,liu2019linearization,ota_2019_wang}.  Instead of using a feedback path, OTA methods utilize an observation receiver to acquire the PA output signal over the channel for DPD optimization. Such methods achieve promising linearization performance and cost savings compared with methods using the feedback path.

Several parametric models for DPD have been utilized, e.g., Volterra-series-based \cite{GMP_2006,liu2014novel,
wang2016low,guan2017digital,beltagy2018direct,hausmair2018modeling,liu2019linearization,ota_2019_wang} or \ac{nn}-based models~\cite{liu2004dynamic,isaksson2005radial,rawat2009adaptive,wu2020residual}. To optimize the models, the works~\cite{hausmair2018modeling,liu2019linearization,ota_2019_wang,GMP_2006,liu2004dynamic,isaksson2005radial,rawat2009adaptive,wu2020residual} consider a sample-based criterion that minimizes the sample error between oversampled signals. This is optimal when the aim is to minimize the \ac{acpr} and \ac{nmse}. When the aim is to minimize the \ac{ser}, however, this approach is not optimal, and a symbol-based criterion is more appropriate.

A symbol-based criterion has been used to optimize the constellation mapper and demapper in~\cite{o2017introduction} and the pulse-shaping/matched filters in~\cite{karanov2018end,aoudia2021waveform} in an end-to-end manner via \ac{sl}, but \ac{sl} requires differential models for all hardware and channel, which limits the usage in real communication systems. Instead, \acf{rl}-based optimization of the constellation mapper and demapper has been proposed in~\cite{n2n_2018,N2N_Jakob_Trans_2019}, which does not require any models for hardware and channel. However, in~\cite{n2n_2018,N2N_Jakob_Trans_2019}, the (de)mapper operates at the same symbol level as the optimization criterion, and no memory effects of hardware are considered. It requires a generalization for \ac{rl}-based DPD optimization with a symbol-based criterion due to the data rate difference between DPD parameters and criterion, and memory effects in the \ac{pa}. To the best of our knowledge, \ac{ota} DPD optimization using \ac{rl} with a symbol-based criterion has not yet been addressed. 
 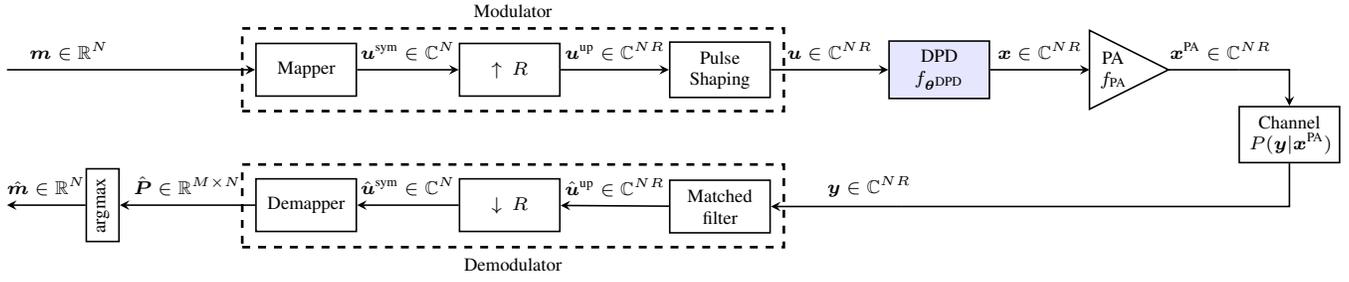
\begin{figure*}[t]
\centering
\begin{tikzpicture}[font=\scriptsize, >=stealth,nd/.style={draw,fill=blue!10,circle,inner sep=0pt,minimum size=5pt}, blk/.style={draw,fill=blue!0,minimum height=0.7cm,text width=1.1cm, text centered}, x=0.6cm,y=0.6cm]
\tikzset{amplifier/.pic={
\draw [fill=blue!0](0,2.5)--(5,0)--(0,-2.5)--cycle node at (1.5,0) {\begin{tabular}{c} PA \\ $f_{\text{PA}}$ \end{tabular}};}}

\path		
		(-6,0)coordinate[](message){} 
	node(const)[blk, right=2.4 of message]{Mapper}  
	node(up)[blk, right=2.25 of const]{$\uparrow R$}
	node(pulse)[blk, right=2.4 of up]{Pulse\\Shaping} 
		node(dpd)[blk,fill=blue!10, right=2.6 of pulse]{DPD \\ $f_{\boldsymbol{\theta}^{\text{DPD}}}$}
		coordinate[right=.5 of dpd](pa_in){}	
coordinate[right=2.2 of dpd](pa)
	(pa)pic[scale=0.35,outer sep =0pt] {amplifier}
	coordinate[right=1.7 of pa](pa_out){}
		coordinate[right=1.6 of pa_out](tx){}	
		node(channel)[blk, below right=.8 and 0 of tx]{Channel\\
		$P(\boldsymbol{y}|\boldsymbol{x}^{\text{PA}})$}
		node(mf)[blk, below=1.8 of pulse]{Matched\\
		filter}
		coordinate[right= 2.9 of mf](rx){} 
		node(down)[blk, below=1.8 of up]{$\downarrow R$}		
		node(de-const)[blk, below=1.8 of const]{Demapper}
		coordinate[left=5.5 of de-const](m_hat){} 
		node[blk,fill opacity=0, dashed, line width=1, minimum width=7.2cm,minimum height=1.1cm](mod)at ($(const)!0.5!(pulse)$){} 
		node[blk,fill opacity=0, dashed, line width=1, minimum width=7.2cm,minimum height=1.1cm](demod)at ($(de-const)!0.5!(mf)$){} 
		node[above=0 of mod](){Modulator}
		node[below=0 of demod](){Demodulator}

;
\node(argmax)[draw, blk,minimum height=.5 cm, minimum width=1, text width=0.2cm,text centered, left=3. of de-const]
{\rotatebox{90}{argmax}};
\coordinate[left=5.5 of const](m){};
\draw[->] (m)--node[above,near start]{$\boldsymbol{m}\in \mathbb{R}^{N}$}(const);
\draw[->] (const)--node[above]{$\boldsymbol{u}^\text{sym} \in \mathbb{C}^{N}$}(up);
\draw[->] (up)--node[above]{$\boldsymbol{u}^\text{up}\in \mathbb{C}^{NR}$}(pulse);
\draw[->] (pulse)--node[above]{$\boldsymbol{u}\in \mathbb{C}^{NR}$}(dpd);
\draw[->] (dpd)--node[above]{$\boldsymbol{x}\in \mathbb{C}^{NR}$}(pa);
\draw[-] (pa_out)--node[above, near end]{$\boldsymbol{x}^\text{PA}\in \mathbb{C}^{NR}$}(tx);
\draw[->] (tx)-|(channel);
\draw[-] (channel)|-(rx);
\draw[->] (rx)--node[above, near start]{$\boldsymbol{y}\in \mathbb{C}^{NR}$}(mf);
\draw[->] (mf)--node[above]{$\hat{\boldsymbol{u}}^\text{up} \in \mathbb{C}^{NR}$}(down);
\draw[->] (down)--node[above]{$\hat{\boldsymbol{u}}^\text{sym} \in \mathbb{C}^{N}$}(de-const);
\draw[->] (de-const)--node[above]{$\hat{\boldsymbol{P}}\in \mathbb{R}^{M\times N}$}(argmax);
\draw[->] (argmax)--node[above]{$\hat{\boldsymbol{m}}\in \mathbb{R}^{N}$}(m_hat);

\end{tikzpicture}
\caption{System model of different blocks in a communication system with an unknown channel, where the block \ac{dpd} linearizes the nonlinearity of the \ac{pa}. The constellation mapping, upsampling, and pulse shaping are referred to as \textit{modulator}. The constellation demapping, downsampling, and matched filtering are referred to as \textit{demodulator}.}
\label{fig:sys_model}
\vspace{-\baselineskip}
\end{figure*} 

In this work, we propose a symbol-based \ac{dpd} optimization algorithm using \ac{ota} observations. Instead of using a sample-based criterion that minimizes the error between oversampled signals, the proposed algorithm minimizes the cross-entropy between transmitted and received symbols. Using the policy gradient theorem~\cite{sutton2018reinforcement}, we generalize the work~\cite{n2n_2018} by connecting the symbol-based criterion with the sample-based policy optimization, which allows to optimize DPD parameters at the sample level. For a \ac{gmp} \ac{pa} and \ac{awgn} channel, we show that the proposed symbol-based \ac{dpd} optimization algorithm achieves \acf{ser} gains over the \ac{ila}-based \ac{dpd} optimization algorithm~\cite{eun1997new} even with full-rate \acp{adc}. Error spectrum results show that, although the DPD optimized by the symbol-based criterion focuses more on reducing the in-band errors (i.e., symbol errors),  out-of-band errors are still at a satisfactory level. 
\subsubsection*{Notation}
Lowercase and uppercase boldface letters denote column vectors and matrices such as $\boldsymbol{x}$ and $\boldsymbol{X}$.  $\mathbb{R},\mathbb{R}_{\geq 0}$, and $\mathbb{C}$ denote real,  non-negative-real, and complex numbers, respectively. $x_n$ or $[\boldsymbol{x}]_{n}$  denote the $n$-th element of $\boldsymbol{x}$, and $\boldsymbol{x}_{n:n+k}$ denotes a vector consisting of the $n$-th to $(n+k)$-th elements of $\boldsymbol{x}$. $\mathbb{E}_{\boldsymbol{x}}\{\cdot\}$ denotes the expectation operator taken over $\boldsymbol{x}$.

\section{System model}

\subsection{System Model}

We consider the communication system  shown in Fig.~\ref{fig:sys_model}. Let  $\boldsymbol{m} \in \mathbb{R}^{N}$ be a message sequence of length $N$, generated from a message set $\mathbb{M}=\{1,...,M\}$. A message sequence $\boldsymbol{m}$ is mapped to a sequence of symbols $\boldsymbol{u}^{\text{sym}}\in \mathbb{C}^{N}$ via a constellation mapping, upsampled with  upsampling rate $R$ to $\boldsymbol{u}^{\text{up}} \in \mathbb{R}^{NR}$, and pulse-shaped to a discrete-time baseband signal $\boldsymbol{u}=[u_1,...,u_n,u_{NR}]^{\text{T}}\in \mathbb{C}^{NR}$, where $u_n$ is the sample transmitted at time instant $n$. To compensate for the nonlinearity of the \ac{pa}, \ac{dpd} is applied to $\boldsymbol{u}$. The DPD is represented by a parametric model $f_{\boldsymbol{\theta}^{\text{DPD}}} : \mathbb{C}^{K_1+1} \rightarrow \mathbb{C}$ with parameters $\boldsymbol{\theta}^{\text{DPD}}$ and input memory length $K_1$. Given the input sequence $\boldsymbol{u}_{n:n-K_1}=[u_n,...,u_{n-K_1}]^{\text{T}}\in \mathbb{C}^{K_1+1}$ to the DPD, the predistorted output $x_n$ can be expressed as
\begin{align}
	 x_n = f_{\boldsymbol{\theta}^{\text{DPD}}}(u_n,...,u_{n-K_1})=f_{\boldsymbol{\theta}^{\text{DPD}}}(\boldsymbol{u}_{n:n-K_1})\,.
	 \label{eq:dpd_in_out}
\end{align}
The predistorted signal $\boldsymbol{x}=[x_1,...,x_n,x_{NR}]^{\text{T}}$ is then amplified by the PA, which is represented by the nonlinear function $f_{\text{PA}} : \mathbb{C}^{K_2+1} \rightarrow \mathbb{C}$ with memory length $K_2$, input $\boldsymbol{x}_{n:n-K_2}=[x_n,...,x_{n-K_2}]^{\text{T}}\in \mathbb{C}^{K_2+1}$, and output $x^{\text{PA}}_n$. The input--output relation of the \ac{pa} can be expressed as
\begin{align}
	 x^{\text{PA}}_{n} = f_{\text{PA}}(x_n,...,x_{n-K_2})=f_{\text{PA}}(\boldsymbol{x}_{n:n-K_2}).
	 \label{eq:pa_in_out}
\end{align}
 The signal $\boldsymbol{x}^{\text{PA}}=[x^{\text{PA}}_1,...,x^{\text{PA}}_n,x^{\text{PA}}_{NR}]^{\text{T}}$ is then sent through a discrete channel with  conditional  distribution $P(\boldsymbol{y}|\boldsymbol{x}^{\text{PA}})$, where $\boldsymbol{y}$ is the channel output. Note that elements in $\boldsymbol{u}_{n:n-K_1}$ and $\boldsymbol{x}_{n:n-K_1}$ with nonpositive indexes are set to zero. Under the assumption of perfect synchronization, the received signal $\boldsymbol{y}$ is demodulated via a matched filter and downsampled with a downsampling rate $R$ to symbols $\hat{\boldsymbol{u}}^{\text{sym}} \in \mathbb{C}^{N}$. Each symbol $\hat{u}^{\text{sym}}_n$ is decoded to a probability vector  $\hat{\boldsymbol{p}}_n \in\mathbb{R}^M_+$ over $M$ messages, where $  \sum_{i=1}^M \hat{p}_i=1$. Finally, the estimated message is obtained by $\hat{m}_n=\arg \max_m [\hat{\boldsymbol{p}}_n]_{m} $. Overall, the estimated message sequence $\hat{\boldsymbol{m}}$ is obtained from the probability matrix $\hat{\boldsymbol{P}}=[\hat{\boldsymbol{p}}_1,...,\hat{\boldsymbol{p}}_N] \in \mathbb{R}_+^{M\times N}$. The demapper can be implemented by an \ac{nn} as in~\cite{o2017introduction,n2n_2018,N2N_Jakob_Trans_2019}, which can be pretrained to have similar decoding performance as the maximum likelihood demapper.
 
 Given a parametric \ac{dpd} model $f_{\boldsymbol{\theta}^{\text{DPD}}}$ with parameters $\boldsymbol{\theta}^{\text{DPD}}$, our objective is to optimize  $\boldsymbol{\theta}^{\text{DPD}}$ with respect to a given loss function $\mathcal{L}$,
 \begin{align}
\hat{\boldsymbol{\theta}}^{\text{DPD}} =\underset{\boldsymbol{\theta}^{\text{DPD}}}{\arg \min} \ \mathcal{L}(\boldsymbol{\theta}^{\text{DPD}}). 
\label{eq:dpd_opt}
\end{align}
 
\subsection{Indirect Learning Architecture}
\begin{figure}[t]
\centering
\begin{tikzpicture}[font=\scriptsize, >=stealth,nd/.style={draw,fill=blue!0,circle,inner sep=0pt,minimum size=5pt}, blk/.style={draw,fill=blue!0,minimum height=0.7cm,text width=1.1cm, text centered}, x=0.6cm,y=0.5cm]

\tikzset{amplifier/.pic={
\draw [fill=blue!0](0,2.5)--(5,0)--(0,-2.5)--cycle node at (1.5,0) {\begin{tabular}{c} PA \\ $f_{\text{PA}}$ \end{tabular}};}}
\tikzset{adc/.pic={
\draw [fill=blue!0](0,2)--(4,2)--(6,0)--(4,-2.)--(0,-2.)--cycle node at (2.5,0) {\begin{tabular}{c} ADC \\ $F_s^{\text{ADC}}$ \end{tabular}};}}

\path		
		coordinate[](dpd_in){} 
		node(dpd)[blk,fill=blue!10,text width=1.3cm, right=1 of dpd_in]{Predistorter\\$f_{\boldsymbol{\theta}^\text{DPD}}$}
		coordinate[right=.5 of dpd](pa_in){}
		node(opt_operation)[circle,draw, minimum size=0.5, below=1.5 of pa_in]{}	
	coordinate[right=2 of dpd](pa)
	(pa)pic[scale=0.35,outer sep =0pt] {amplifier}
	coordinate[right=1.7 of pa](pa_out0){}
		coordinate[right=5.3 of pa](pa_out1){}
		coordinate[right=5 of pa_out0](tx){}	
		node(post_d)[blk,fill=blue!10, text width=1.3cm, below left=2.5 and -0.1 of pa_out0]{Postdistorter\\$f_{\boldsymbol{\theta}^\text{DPD}}$}
		coordinate[right=3 of opt_operation](ILA_loss1){}
		coordinate[below left=0.5 and 0.1 of post_d](ILA_loss2){}
		coordinate[below left=0.5 and 0. of dpd](ILA_loss3){}
		coordinate[above right=0.5 and 0.1 of dpd](ILA_loss4){}
		coordinate[right=1. of post_d](adc)
		(adc)pic[scale=0.35,outer sep =0pt] {adc}
		coordinate[right=2.0 of adc](adc_in)
		node[blk,fill opacity=0, dashed, line width=1, minimum width=2.65cm,minimum height=1.2cm](mod)at (10.5,-3.2){} 

;
\draw[->] (dpd_in)--node[above]{$\boldsymbol{u}$}(dpd);
\draw[->] (dpd)--node[above,near start]{$\boldsymbol{x}$}(pa);
\draw[->] (pa_out0)--node[above,pos=0.4]{$\boldsymbol{x}_\text{PA}$}(tx);
\draw[->] (pa_out1)|-node[near end,right=-0.15]{\begin{tabular}{c} Feedback \\ path \end{tabular}}(adc_in);
\draw[->] (pa_in)--node[above]{}(opt_operation);
\draw[<-] (opt_operation)|-node[left,near start]{$\hat{\boldsymbol{x}}$}(post_d);
\draw[-] (opt_operation)--node[above,near start]{$L(\boldsymbol{\theta}^{\text{DPD}})$}(ILA_loss1);
\draw[->] (ILA_loss1)--(ILA_loss2);
\draw[dotted] (ILA_loss2)-|node[above,near start]{Copy parameters}(ILA_loss3);
\draw[->] (ILA_loss3)--(ILA_loss4);
\draw[->] (adc)--node[above]{$\hat{\boldsymbol{x}}_{\text{PA}}$}(post_d);

\path
	node(post_d)[blk,fill=blue!10, text width=1.3cm, below left=2.5 and -0.1 of pa_out0]{Postdistorter\\$f_{\boldsymbol{\theta}^\text{DPD}}$}
	node(dpd)[blk,fill=blue!10,text width=1.3cm, right=1 of dpd_in]{Predistorter\\$f_{\boldsymbol{\theta}^\text{DPD}}$}
;

\end{tikzpicture}
\caption{Block diagram of the ILA. Learning the inverse behavior of the PA by minimizing waveform errors, a postdistorter is optimized, which is then utilized as the predistorter.}
\label{fig:sys_model_ILA}
\end{figure}
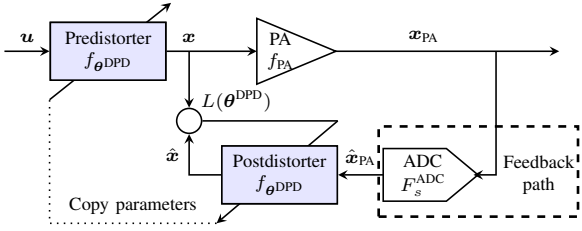

The \ac{ila}~\cite{eun1997new} indirectly optimizes the \ac{dpd} parameters by estimating an inverse behavior of the \ac{pa} as shown in Fig.~\ref{fig:sys_model_ILA}.  The optimization of the parameters of the distorter is conducted after the PA, and hence the distorter is referred to as the \textit{postdistorter}. After the parameter optimization, the postdistorter is used as the \textit{predistorter}, placed before the PA. To optimize the postdistorter, a feedback path is required for signal acquisition of the PA. Here we consider an \ac{adc} in the feedback path with a sampling rate $F_s^{\text{ADC}}$. We denote the output of the ADC by $\hat{\boldsymbol{x}}^{\text{PA}}$. The parameters of the postdistorter are usually optimized by minimizing the \ac{mse} between the postdistorter output signal $\hat{\boldsymbol{x}}$ and the PA input signal $\boldsymbol{x}$. In this case, the  loss function $\mathcal{L}$ can be expressed as
\begin{align}
	\mathcal{L}(\boldsymbol{\theta}^{\text{DPD}}) &=\mathbb{E}_{\boldsymbol{x}}\left\{ |x_n- \hat{x}_{n}|^2 \right\} \nonumber\\
	&= \mathbb{E}_{\boldsymbol{x}}\left\{ |x_n - f_{\boldsymbol{\theta}^{\text{DPD}}}(\hat{\boldsymbol{x}}^{\text{PA}}_{n:n-K_1}))|^2 \right \},
	\label{eq:loss_mse}
\end{align}
where $\hat{\boldsymbol{x}}^{\text{PA}}_{n:n-K_1}=[\hat{x}^{\text{PA}}_n,...,\hat{x}^{\text{PA}}_{n-K_1}]^{\text{T}}$ is the input of the postdistorter with memory $K_1$.
Substituting~\eqref{eq:loss_mse} into~\eqref{eq:dpd_opt}, $\boldsymbol{\theta}^{\text{DPD}}$ is optimized by minimizing the sample difference between the postdistoter output $\hat{\boldsymbol{x}}$ and the  PA output $\hat{\boldsymbol{x}}^{\text{PA}}$  collected by the feedback path, i.e., the optimization is based on a \textit{sample-based} criterion. 
\section{Proposed Symbol-Based DPD \\Optimization Algorithm}

We propose to optimize the DPD based on a \textit{symbol-based} criterion using \ac{ota} measurements. Specifically, the symbol-based criterion minimizes the cross-entropy between the transmitted and received messages $\boldsymbol{m}$ and $\hat{\boldsymbol{m}}$.

\subsection{Symbol-Based Criterion}
To implement the symbol-based criterion for DPD optimization, we consider a symbol-based cross-entropy loss function, which defines the transmitted message sequence $\boldsymbol{m}$ and received message probabilities $\hat{\boldsymbol{P}}$ as
\begin{align}
	\mathcal{L}(\boldsymbol{\theta}^{\text{DPD}})  =- \mathbb{E}_{\boldsymbol{m}} \{ \underbrace{ \log([\hat{\boldsymbol{p}}_n]_{m_n})}_{l_n^{\text{CE}}} \},
	\label{eq:mes_ce_loss}
\end{align}
where the loss function between the transmitted message $m_n$ and the corresponding vector of probabilities   $\hat{\boldsymbol{p}}_n$ is denoted by $l_n^{\text{CE}}$, referred to as the cross-entropy per-example loss.

\subsection{Supervised Learning}
Assuming that all blocks in the communication system are differentiable, the derivatives of the loss function at the receiver side with respect to any trainable parameter in the system can be analytically calculated using the chain rule. Thus, $\boldsymbol{\theta}^{\text{DPD}}$ can be updated through back-propagation via a mini-batch \ac{sgd} algorithm as
\begin{align}
	\boldsymbol{\theta}^{\text{DPD}}_{j+1} =  \boldsymbol{\theta}^{\text{DPD}}_{j} - \eta \nabla_{\boldsymbol{\theta}^{\text{DPD}}}\mathcal{L}(\boldsymbol{\theta}^{\text{DPD}}_{j}),
	\label{eq:sgd_theta_DPD}
\end{align}
where $\eta>0$ denotes the learning rate and $\nabla_{\boldsymbol{\theta}^{\text{DPD}}}\mathcal{L}(\boldsymbol{\theta}^{\text{DPD}}_{j})$ is the derivative of $\mathcal{L}$ with respect to $\boldsymbol{\theta}^{\text{DPD}}_j$ at step $j$. 

However, in a real communication system, most of the blocks are non-differentiable, and thus it is infeasible to apply the chain rule to calculate $\nabla_{\boldsymbol{\theta}^{\text{DPD}}}\mathcal{L}$. Although we can circumvent this problem using surrogate parametric models of the hardware components, e.g., pretrained PA model in~\cite{paaso2008comparison}, it is cumbersome to pretrain such models, and the performance  highly depends on the model accuracy. 
\subsection{Reinforcement Learning}
\Ac{rl} is defined as a learning process of how an \textit{agent} takes \textit{actions} in an environment to minimize a given loss~\cite{sutton2018reinforcement}. The optimization of the \ac{dpd} parameters can be viewed through the lens of a \ac{rl} problem. The \ac{dpd} acts as an agent that takes actions following a \textit{policy}, which is optimized to minimize the loss $\mathcal{L}$. RL has already been used in the context of communications, e.g.,  to optimize the constellation mapper and demapper~\cite{n2n_2018,N2N_Jakob_Trans_2019}. Here, we consider a different scenario  with more components, which raises the problem of how to optimize the DPD parameters $\boldsymbol{\theta}^{\text{DPD}}$ at the sample-based using a loss function $\mathcal{L}$ at the symbol-based.

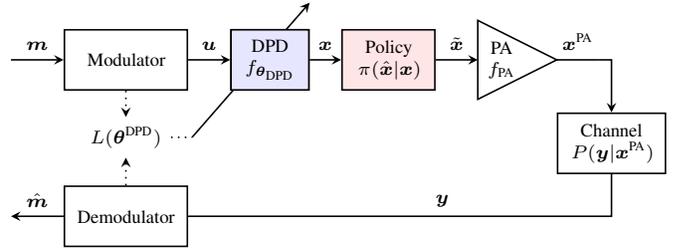
\begin{figure}[t]
\centering
\begin{tikzpicture}[font=\scriptsize, >=stealth,nd/.style={draw,fill=blue!0,circle,inner sep=0pt,minimum size=5pt}, blk/.style={draw,fill=blue!0,minimum height=0.8cm,text width=1.1cm, text centered}, x=0.7cm,y=0.7cm]

\tikzset{amplifier/.pic={
\draw [fill=blue!0](0,2.5)--(5,0)--(0,-2.5)--cycle node at (1.5,0) {\begin{tabular}{c} PA \\ $f_{\text{PA}}$ \end{tabular}};}}

\path		
		(-2,0)coordinate[](message){} 
	node(mod)[blk, right=1 of message, text width=1.4cm]{Modulator}
	node(dpd)[blk,fill=blue!10, right=0.8 of mod,text width=0.8cm]{DPD\\$f_{\boldsymbol{\theta}_ \text{DPD}}$}
		coordinate[right=.8 of dpd](pa_in){}	
		node(pertub)[blk,fill=red!10, right=0.6 of dpd,text width=1.0cm]{Policy\\$\pi(\hat{\boldsymbol{x}}|\boldsymbol{x})$}
		coordinate[right=0.8 of pertub](pa)
	(pa)pic[scale=0.3,outer sep =0pt] {amplifier}
	coordinate[right=1.5 of pa](pa_out){}

		coordinate[right=.8 of pa_out](tx){}	
		node(channel)[blk, below right=1. and -0.8 of tx,text width=1.2cm]{Channel\\
		$P(\boldsymbol{y}|\boldsymbol{x}^{\text{PA}})$}
		coordinate[below= 2.6 of tx](rx){} 
		node(demod)[blk, below =1.8 of mod,text width=1.4cm]{Demodulator}
		coordinate[left=1 of demod](m_hat){} 
		node(loss)[below=0.5 of mod]{$L(\boldsymbol{\theta}^{\text{DPD}})$}
		coordinate[right=.4 of loss](loss_2){} 
		coordinate[above right=0.5 and -0 of dpd](loss_3){} 
;

\draw[->] (message)--node[above]{$\boldsymbol{m}$}(mod);
\draw[->] (mod)--node[above]{$\boldsymbol{u}$}(dpd);
\draw[->] (dpd)--node[above]{$\boldsymbol{x}$}(pertub);
\draw[->] (pertub)--node[above]{$\tilde{\boldsymbol{x}}$}(pa);
\draw[-] (pa_out)--node[above]{$\boldsymbol{x}^\text{PA}$}(tx);
\draw[->] (tx)-|(channel);
\draw[-] (channel)|-node[above, pos=0.7]{$\boldsymbol{y}$}(demod);
\draw[->] (demod)--node[above]{$\hat{\boldsymbol{m}}$}(m_hat);
\draw[dotted,->] (demod)--(loss);
\draw[dotted,->] (mod)--(loss);
\draw[dotted] (loss)--(loss_2);
\draw[->] (loss_2)--(loss_3);

\path
	node(dpd2)[blk,fill=blue!10, right=0.8 of mod,text width=0.8cm]{DPD\\$f_{\boldsymbol{\theta}_ \text{DPD}}$}
;
\end{tikzpicture}
\caption{Block diagram of the proposed symbol-based optimization for DPD. The \ac{dpd} parameters $\boldsymbol{\theta}^{\text{DPD}}$ are optimized using \ac{ota} observations at the receiver side by minimizing a symbol-based cross-entropy between symbols.}
\label{fig:sys_model_N2N}
\vspace{-\baselineskip}
\end{figure}

As shown in Fig.~\ref{fig:sys_model_N2N}, we consider a Gaussian policy $\pi(\tilde{\boldsymbol{x}}|\boldsymbol{x})$ that converts the deterministic actions $\boldsymbol{x}$ to stochastic actions $\tilde{\boldsymbol{x}}$, which enables the \textit{exploration} of possible actions. For an arbitrary action $x_n$, we consider an independent Gaussian policy $\pi(\tilde{x}_n|x_n)$, which generates output $\tilde{x}_n$ by adding a Gaussian perturbation $w \sim \mathcal{C N}(0,\sigma_{\pi}^2)$ as 
\begin{align}
	\tilde{x}_n = \sqrt{1-\sigma^2_{\pi}}x_n + w,
	\label{eq:add_perturbation}
\end{align}
where $\sigma_{\pi}^2$ is the variance of  the perturbation, which is a  hyper-parameter that is fixed during the training. Thus, the policy $\pi(\tilde{x}_n|x_n)$ is the \ac{pdf} of a complex Gaussian variable with mean $\sqrt{1-\sigma^2_{\pi}}x_n = \sqrt{1-\sigma^2_{\pi}}f_{\boldsymbol{\theta}^{\text{DPD}}}(\boldsymbol{u}_{n:n-K_1})$ and variance $\sigma_{\pi}^2$,
\begin{align}
	\pi(\tilde{x}_n|x_n) \propto  \exp \left(-\frac{\left|\tilde{x}_n - \sqrt{1-\sigma_{\pi}^2} f_{\boldsymbol{\theta}^{\text{DPD}}}(\boldsymbol{u}_{n:n-K_1})\right|^2}{\sigma_{\pi}^{2}} \right).
	\label{eq:Gaus_pdf}
\end{align} 

Based on the received observation $y_n$, the receiver can compute the corresponding per-example loss $l_n$ by $l_n^{\text{CE}}$. The loss $l_n$ is related to a subset of the entire sequence $\boldsymbol{x}$ because of the memory effects of the PA and convolution operation in the matched filtering. Denote this subset by $\boldsymbol{x}_{(G)}=\{x_{nR-G},...,x_{nR+G}\}$, where $G$ denotes the number of signals being related. Because of the convolution operation in the pulse shaping, the subset $\boldsymbol{x}_{(G)}$ depends on a subset of the messages in $\boldsymbol{m}$, denoted by $\boldsymbol{m}_{(G)}=\{m_{nR-G},...,m_{nR+G}\}$. Similarly, we can define $\tilde{\boldsymbol{x}}_{(G)}=\{\tilde{x}_{nR-G},...,\tilde{x}_{nR+G}\}$ and $\boldsymbol{y}_{(G)}=\{y_{nR-G},...,y_{nR+G}\}$.

The objective of the DPD agent is to minimize the loss function $\mathcal{L}(\boldsymbol{\theta}^{\text{DPD}})$, defined as
\begin{equation}
\begin{aligned}
	\mathcal{L}(\boldsymbol{\theta}^{\text{DPD}}) 
	&\triangleq \mathbb{E}_{\boldsymbol{m},\tilde{\boldsymbol{x}},\boldsymbol{y}}\left\{ l_n\right \}\,,
\end{aligned}
\end{equation}
where $l_n$ is $l_n^{\text{CE}}$ (see \eqref{eq:mes_ce_loss}). In order to minimize the loss function, we compute the gradient with respect to $\boldsymbol{\theta}^{\text{DPD}}$ according to the following proposition.

\begin{proposition}
\label{prop:prop}
The gradient of $\mathcal{L}(\boldsymbol{\theta}^{\text{DPD}})$ with respect to $\boldsymbol{\theta}^{\text{DPD}}$ is approximated by
\begin{align}
	&\nabla_{\boldsymbol{\theta}^{\text{DPD}}} \mathcal{L}(\boldsymbol{\theta}^{\text{DPD}})\label{eq:deriv_J_theta_before}\\ 
	&\approx  \frac{2\sqrt{1-\sigma_{\pi}^2}}{N\sigma_{\pi}^2}\sum_{n=1}^{N} l_{n} \sum_{g=-G}^{G} \left(\tilde{x}_n - \sqrt{1-\sigma_{\pi}^2} f_{\boldsymbol{\theta}^{\text{DPD}}}(\boldsymbol{u}_{n:n-K_1})\right)\notag \\
	&\times \nabla_{\boldsymbol{\theta}^{\text{DPD}}} f_{\boldsymbol{\theta}^{\text{DPD}}}(\boldsymbol{u}_{n:n-K_1})\,. \notag
	%
\end{align}
\end{proposition}

\begin{IEEEproof}
Exploiting the policy gradient theorem~\cite{sutton2018reinforcement} and using the fact that  $\nabla \log(\pi) = (\nabla \pi)/\pi$, we can write
\begin{align}
  	\nabla_{\boldsymbol{\theta}^{\text{DPD}}} \mathcal{L}(\boldsymbol{\theta}^{\text{DPD}})= \mathbb{E}_{\boldsymbol{m},\tilde{\boldsymbol{x}},\boldsymbol{y}}\left\{  l_n \  \nabla_{\boldsymbol{\theta}^{\text{DPD}}} \log\left(\pi\left(\tilde{\boldsymbol{x}}| \boldsymbol{x}\right)\right)\right \}\,.
  	\label{eq:Prop1}
\end{align}
The loss $l_n$ is related to a fraction of $\boldsymbol{x}$ consisting of $G$ signals. Restricting \eqref{eq:Prop1} to this subset of signals, we can approximate it as 
\begin{align}
  &\mathbb{E}_{\boldsymbol{m},\tilde{\boldsymbol{x}},\boldsymbol{y}}\left\{  l_n \  \nabla_{\boldsymbol{\theta}^{\text{DPD}}} \log\left(\pi\left(\tilde{\boldsymbol{x}}_{(G)}| \boldsymbol{x}_{(G)}\right)\right)\right \}\label{eq:Prop2}\\
  &=\mathbb{E}_{\boldsymbol{m},\tilde{\boldsymbol{x}},\boldsymbol{y}}\Big\{   l_n \ \nabla_{\boldsymbol{\theta}^{\text{DPD}}} \log\Big( \prod_{g=-G}^{G} \pi(\tilde{x}_{nR+g}| x_{nR+g})\Big)\Big\}, \nonumber
\end{align}
where \eqref{eq:Prop2} follows as the  conditional probability $\pi(\tilde{\boldsymbol{x}}_{(G)}| \boldsymbol{x}_{(G)})$ reduces to the product of  conditional probabilities $\pi(\tilde{x}_{nR+g}|x_{nR+g})$ due to the independence of each action. Now, using the fact that the product of logarithms can be written as the  logarithm of a sum and approximating the expectation by the average of  $N$ (correlated) samples from the underlying distribution $p(\boldsymbol{m},\tilde{\boldsymbol{x}},\boldsymbol{y})$, we obtain
\begin{align}
  	& \nabla_{\boldsymbol{\theta}^{\text{DPD}}} \mathcal{L}(\boldsymbol{\theta}^{\text{DPD}})  \label{eq:Prop3}\\
  	& \approx \frac{1}{N}\sum_{n=1}^{N} l_{n} \sum_{g=-G}^{G} \nabla_{\boldsymbol{\theta}^{\text{DPD}}}\log(\pi(\tilde{x}_{nR+g}| x_{{nR+g}}))\,.\notag 
\end{align}
From \eqref{eq:Gaus_pdf}, it follows that  
\begin{align}
&\nabla_{\boldsymbol{\theta}^{\text{DPD}}} \log(\pi(\tilde{x}_{nR+g} | x_{nR+g}))\label{eq:nalblapi}\\
&=\frac{2\sqrt{1-\sigma_{\pi}^2}}{\sigma_{\pi}^2}\big(\tilde{x}_n - \sqrt{1-\sigma_{\pi}^2} f_{\boldsymbol{\theta}^{\text{DPD}}}(\boldsymbol{u}_{n:n-K_1})\big)\nabla_{\boldsymbol{\theta}^{\text{DPD}}}f_{\boldsymbol{\theta}^{\text{DPD}}}(\cdot).\nonumber
\end{align}
Substituting \eqref{eq:nalblapi} into \eqref{eq:Prop3} completes the proof.
\end{IEEEproof}

The approximation of the gradient of the loss function $\mathcal{L}$ with respect to  $\boldsymbol{\theta}^{\text{DPD}}$ in Proposition~\ref{prop:prop}  can be used to optimize $\boldsymbol{\theta}^{\text{DPD}}$ via any \ac{sgd}-based algorithm as in~\eqref{eq:sgd_theta_DPD}. 

\begin{algorithm}[t]
\caption{: Symbol-based optimization for $\boldsymbol{\theta}^{\text{DPD}}$.}
 \hspace*{\algorithmicindent} \textbf{Input:} $N$,$N_{\text{B}}$ $\sigma_{\pi}^2$, $R$, $G$
\begin{algorithmic}[1]
\FOR{A number of iterations $N_{\text{B}}$}
    \STATE Messages to symbols: $\boldsymbol{m} \rightarrow \boldsymbol{u}^{\text{sym}}$ 
    \STATE Upsampled and pulse-shaped with an upsampling rate $R$: $\boldsymbol{u}^{\text{sym}} \rightarrow \boldsymbol{u}^{\text{up}} \rightarrow \boldsymbol{u}$ 
    \STATE DPD: $f_{\boldsymbol{\theta}^{\text{DPD}}}(\boldsymbol{u}) \rightarrow \boldsymbol{x}$ via~\eqref{eq:dpd_in_out}
    \STATE Policy: $\pi(\cdot|\boldsymbol{x}) \rightarrow \tilde{\boldsymbol{x}}$ via~\eqref{eq:add_perturbation}
    \STATE PA: $f_{\text{PA}}(\tilde{\boldsymbol{x}}) \rightarrow \boldsymbol{x}^{\text{PA}}$ via~\eqref{eq:pa_in_out}
    \STATE Channel: $ P(\cdot|\boldsymbol{x}^{\text{PA}}) \rightarrow \boldsymbol{y}$ 
    \STATE Matched filtered and downsampled with a downsampling rate $R$: $\boldsymbol{y} \rightarrow \hat{\boldsymbol{u}}^{\text{up}} \rightarrow \hat{\boldsymbol{u}}^{\text{sym}}$
    \STATE Symbols to messages: $\hat{\boldsymbol{u}}^{\text{sym}} \rightarrow \hat{\boldsymbol{m}}$
	\STATE Per-example losses: $l_n$
	\STATE Update $\boldsymbol{\theta}^{\text{DPD}}$: $\text{SGD}(\boldsymbol{\theta}^{\text{DPD}},\mathcal{L}) \rightarrow \boldsymbol{\theta}^{\text{DPD}}$ via~\eqref{eq:sgd_theta_DPD} and \eqref{eq:deriv_J_theta_before}
\ENDFOR
\STATE Remove policy $\pi(\cdot)$
\end{algorithmic}
  \label{alg:dpd_training}
 \end{algorithm}

The details of the symbol-based DPD optimization procedure of $\boldsymbol{\theta}^{\text{DPD}}$ are given in Algorithm~\ref{alg:dpd_training}. First, each batch of $N$ messages, $\boldsymbol{m}$, is generated and transformed to a sequence of symbols $\boldsymbol{u}^{\text{sym}}$, then upsampled and pulse-shaped to $\boldsymbol{u}$ with an upsampling rate $R$. After the DPD model $f_{\boldsymbol{\theta}^{\text{DPD}}}(\cdot)$, the Gaussian policy is applied
by adding a perturbation $w$ to generate exploration samples $\tilde{\boldsymbol{x}}$ as in~\eqref{eq:add_perturbation}. These samples are
 sent through the PA and channel. The output of the channel,  $\boldsymbol{y}$, is transformed to  $\hat{\boldsymbol{u}}^{\text{sym}}$ by the matched filter,  downsampled with a downsampling rate $R$, and eventually decoded to  $\hat{\boldsymbol{m}}$. Then, the cross-entropy per-example losses, $l_n^{\text{CE}}$, are calculated, which are assumed to be known at the transmitter via a reliable feedback channel~\cite{n2n_2018,Back_no_channel_Raj_2018,N2N_Jakob_Trans_2019}. Finally, $\boldsymbol{\theta}^{\text{DPD}}$ is updated by an \ac{sgd}-based algorithm as in~\eqref{eq:sgd_theta_DPD}, where  $\nabla_{\boldsymbol{\theta}^{\text{DPD}}} \mathcal{L}(\boldsymbol{\theta}^{\text{DPD}})$ is computed using~\eqref{eq:deriv_J_theta_before}. The whole procedure is iterated for a number of iterations $N_{\text{B}}$, and the policy $\pi(\cdot)$ is removed.

\addtolength{\topmargin}{0.01in}
\begin{table}[t]
\centering
\caption{Parameter setup}
\label{tab:coef_set}
\begin{tabular}{@{}cccccccc@{}}
\toprule
$M$  & $R$  & $N$ & $G$   & $\sigma_{\pi}^2$ & $\sigma_{\text{ch}}$ [V]  
\\ \midrule
$64$ & $4$ & $1024$ & $3$ & $0.08$ & $0.3$ 
\\ \bottomrule
\end{tabular}
\end{table}
\begin{figure}[!t]
	\vspace*{-0.5\baselineskip}
	\centering
	\begin{tikzpicture}[every node/.style={minimum size=1cm},font=\scriptsize, >=stealth,nd/.style={draw,fill=blue!10,circle,inner sep=0pt},blk/.style={draw,fill=blue!10,minimum height=2. cm, minimum width=0.5 cm, text centered}, x=0.6cm,y=0.5cm]

\coordinate[](u) at (0,0);
  \node(c2r)[draw, blk,fill=blue!0, right=1 of u] {\rotatebox{90}{$\mathbb{C}2\mathbb{R}$}};
  \coordinate[right=0.25 of c2r](residual_1){};
  \node(dense_1)[draw, blk, right=0.6 of c2r] {\rotatebox{90}{Dense(12, ReLu)}};
    \coordinate[below=2.5 of residual_1](residual_2);

    \node(dense_2)[draw, blk, right=0.5 of dense_1] {\rotatebox{90}{Dense(12, ReLu)}};
   \node(dense_3)[draw, blk, right=0.5 of dense_2] {\rotatebox{90}{Dense(2, Linear)}};
   \node(norm)[draw,blk, fill=blue!0,  right=0.6 of dense_3] {\rotatebox{90}{Normalization}};
    \coordinate[right=0.25 of dense_3](residual_3){};
    \coordinate[below=2.5 of residual_3](residual_4){};
   \node(perturb)[draw, blk,fill=red!10,minimum height=0.8cm, minimum width=.5 cm, text width=0.6cm, right=0.8 of norm] {Policy \\$\pi(\cdot)$};
   \node(r2c)[draw, blk, fill=blue!0, right=0.5 of perturb] {\rotatebox{90}{$\mathbb{R}2\mathbb{C}$}};
     \node[blk, fill opacity=0.0, dashed, line width=1, minimum width=4.3cm,minimum height=2.2cm](arden)at ($(c2r)!0.5!(norm)$){};
     \node[above=-0.3cm of arden](arden_name){$f_{\boldsymbol{\theta}^{\text{DPD}}}$};
     \coordinate[right=0.6 of r2c](x_p){};

    \draw [->] (u)--node[above=-0.3cm]{$\boldsymbol{u}$}(c2r);
    \draw [->] (c2r)--(dense_1);
    \draw [->] (dense_1)--(dense_2);
    \draw [->] (dense_2)--(dense_3);
    \draw [->] (dense_3)--(norm);
    \draw [->] (norm)--node[above=-0.3cm, near end]{$\boldsymbol{x}$}(perturb);
    \draw [->] (perturb)--(r2c);
    \draw [-] (residual_1)--(residual_2);
    \draw [-] (residual_2)--node[below=-0.4cm]{Residual Connection}(residual_4);
    \draw [->] (residual_4)--(residual_3);
    \draw [->] (r2c)--node[above=-0.3cm]{$\tilde{\boldsymbol{x}}$}(x_p);

\end{tikzpicture}
	\vspace*{-0.5\baselineskip}
	\caption{Structure of the DPD model  $f_{\boldsymbol{\theta}^{\text{DPD}}}$ using R2TDNN~\cite{wu2020residual}, which is placed before the policy block. The normalization block aligns the average output power of DPD to be the same as its input.}
	\label{fig:r2tdnn}
\end{figure}
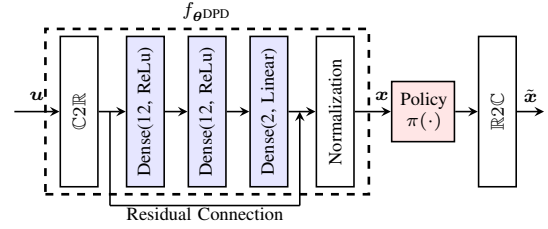
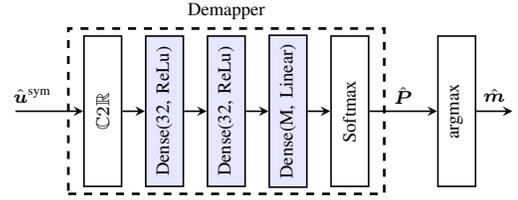
\begin{figure}[!t]
\centering
	\vspace*{-1\baselineskip}
	\begin{tikzpicture}[every node/.style={minimum size=1cm},font=\scriptsize, >=stealth,nd/.style={draw,fill=blue!10,circle,inner sep=0pt},blk/.style={draw,fill=blue!10,minimum height=2 cm, minimum width=0.5 cm, text centered}, x=0.6cm,y=0.5cm]

\coordinate[](u_sym) at (0,0);
  \node(c2r)[draw, blk, fill=blue!0, right=1.5 of u_sym] {\rotatebox{90}{$\mathbb{C}2\mathbb{R}$}};
   \node(dense_1)[draw, blk, right=0.5 of c2r] {\rotatebox{90}{Dense(32, ReLu)}};
   \node(dense_2)[draw, blk, right=0.5 of dense_1] {\rotatebox{90}{Dense(32, ReLu)}};
   \node(dense_3)[draw, blk, right=0.5 of dense_2] {\rotatebox{90}{Dense(M, Linear)}};
   \node(softmax)[draw, blk, fill=blue!0, right=0.5 of dense_3] {\rotatebox{90}{Softmax}};
   \node(argmax)[draw, blk, fill=blue!0, right=1.5 of softmax] {\rotatebox{90}{argmax}};
   \coordinate[right=0.8 of argmax](m_hat);
	\node[blk, fill opacity=0.0, dashed, line width=1, minimum width=4.2cm,minimum height=2.2cm](f_D)at ($(c2r)!0.5!(softmax)$){};
   \node[above=-0.3 cm of f_D]{Demapper};

   \draw [->] (u_sym)--node[above=-0.3cm, near start]{$\hat{\boldsymbol{u}}^{\text{sym}}$}(c2r);
      \draw [->] (c2r)--(dense_1);
   \draw [->] (dense_1)--(dense_2);
   \draw [->] (dense_2)--(dense_3);
   \draw [->] (dense_3)--(softmax);
   \draw [->] (softmax)--node[above=-0.3cm]{$\hat{\boldsymbol{P}}$}(argmax);
   \draw [->] (argmax)--node[above=-0.3cm]{$\hat{\boldsymbol{m}}$}(m_hat);

\end{tikzpicture}
	\caption{Structure of the NN-based $64$ QAM decoder, whose parameters are pretrained and frozen during the training of DPD.}
	\label{fig:f_Decod}
\end{figure}

\section{Numerical Results}

\subsection{Setup}
\subsubsection{Parameters}
We consider a \ac{gmp} model as the PA with nonlinear orders $K_a=K_b=K_c=7$, memory lengths $L_a=L_b=Lc=3$, and cross-term lengths $M_b=M_c=1$~\cite[Eq.~(23)]{GMP_2006}. The corresponding parameters are estimated from the measurements of the RF WebLab\footnote{The RF WebLab is a PA measurement setup that can be remotely accessed at \url{www.dpdcompetition.com}.} using the ILA and a $55$ MHz signal. The measured saturation point and measurement noise standard deviation of the PA are $20.9$ V ($\approx 36.4$ dBm) and $0.053$ V. We consider a $50$ $\Omega$ load impedance.  The remaining parameters are given in Table~\ref{tab:coef_set}. The number of training messages in $\boldsymbol{m}$ for each batch is $N=1024$. The optimizer for gradient descent is Adam~\cite{kingma2014adam} with a learning rate $0.001$. The messages are mapped to a sequence of symbols $\boldsymbol{u}^{\text{sym}}$ according to a $M=64$ \ac{qam} constellation. The sampling rate is $200$ MHz, with upsampling and downsampling rate $R=4$. The pulse shaping filter is a \ac{rrc} filter with a roll-off factor $0.1$, so the bandwidth of the baseband signal $\boldsymbol{u}$ is $55$ MHz. The \ac{papr} of the signal is $10.3$ dB, which is similar to that of an \ac{ofdm} signal.  We set $G=3$ in~\eqref{eq:deriv_J_theta_before}, and the perturbation variance $\sigma^2_{\pi}=0.08$. We consider an \ac{awgn} channel with  fixed noise standard deviation $\sigma_{\text{ch}}=0.3$ Volt. We consider a simulated ADC in the feedback path of the ILA with infinite resolution but three different sampling rates $F_s^{\text{ADC}}=110$ MHz, $220$ MHz, and $550$ MHz (referred to full-rate ADC). Note that $F_s^{\text{ADC}}$ needs to be larger than the Nyquist rate ($110$ MHz) of the distortion-free $55$ MHz signal to capture the out-of-band behavior of the PA output signal.  
\subsubsection{Model Structures}
For the DPD model, we choose both the \ac{gmp}~\cite{GMP_2006} and the \ac{r2tdnn}, where the latter is from our previous work~\cite{wu2020residual} and shows to outperform many other NNs and \ac{gmp} for DPD in terms of complexity versus performance. We consider the same parameter settings (i.e., nonlinear order, memory length, and cross-term length) for the GMP DPD. The specific structure of \ac{r2tdnn} along with policy $\pi(\cdot)$ is shown in~Fig.~\ref{fig:r2tdnn}.  The block $\mathbb{C}2\mathbb{R}$ transforms the complex-valued signal, $\boldsymbol{u}$, to a real-valued signal. We consider two hidden layers with $12$ neurons each, and $3$ input memory length as the same as the PA model. The output of the linear layer is added with the input via the residual connection and then normalized by the normalization layer, which ensures that the average output power of the DPD is the same as its average input power. 

We consider an \ac{nn}-based $M=64$ QAM decoder as shown in Fig.~\ref{fig:f_Decod}, which are trained to have similar performance as the maximum likelihood decoder. The learned detector is frozen during the training of the DPD. Specifically, the softmax layer outputs a probability vector over $M$ messages, where the largest probability represents the predicted message.

\subsection{Simulation results}
\subsubsection{SER versus average PA output power}
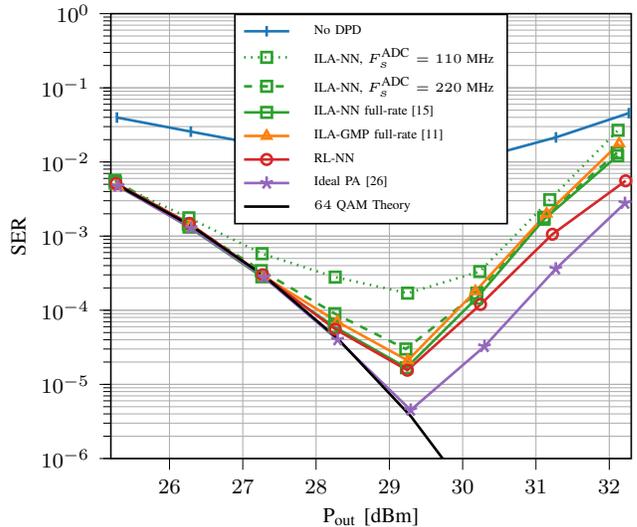
\begin{figure}[t]
	\centering
	\begin{tikzpicture}[font=\footnotesize]

\definecolor{color0}{rgb}{0.12156862745098,0.466666666666667,0.705882352941177}
\definecolor{color1}{rgb}{1,0.498039215686275,0.0549019607843137}
\definecolor{color2}{rgb}{0.172549019607843,0.627450980392157,0.172549019607843}
\definecolor{color3}{rgb}{0.83921568627451,0.152941176470588,0.156862745098039}
\definecolor{color4}{rgb}{0.580392156862745,0.403921568627451,0.741176470588235}

\begin{axis}[
width=8.5cm,
height=7.5cm,
legend cell align={left},
legend style={
  fill opacity=1,
  draw opacity=1,
  text opacity=1,
  at={(0.5,1)},
  anchor=north,
  draw=white!0!black
},
log basis y={10},
tick align=outside,
tick pos=left,
x grid style={white!69.0196078431373!black},
xlabel={P$_{\text{out}}$ [dBm]},
xmajorgrids,
xmin=25.2, xmax=32.3,
xtick style={color=black},
y grid style={white!69.0196078431373!black},
ylabel={SER},
ymajorgrids,
yminorgrids,
ymin=1e-06, ymax=1e0,
ymode=log,
ytick style={color=black}
]
\addplot [semithick, color0, line width=1.0pt, mark=|, mark size=2, mark options={solid}]
table {%
25.2828884124756 0.0397675216197968
26.2913360595703 0.0256675243377686
27.291711807251 0.0167550265789032
28.2864017486572 0.0114450335502625
29.2840270996094 0.00952001810073853
30.2699165344238 0.0115725219249725
31.2727851867676 0.021467524766922
32.267707824707 0.0455775201320648
};
\addlegendentry{\tiny No DPD}

\addplot [semithick,color2, dotted, line width=1.0pt,mark=square, mark size=2, mark options={solid}]
table {%
25.2609462738037 0.00516752600669861
26.2641296386719 0.0017650306224823
27.2626781463623 0.000580024719238281
28.2607765197754 0.000280022621154785
29.2488384246826 0.000170028209686279
30.2335395812988 0.000332528352737427
31.1888637542725 0.00309253349304199
32.1200218200684 0.0267752623558044
};
\addlegendentry{\tiny ILA-NN, $F_s^{\text{ADC}}=110$ MHz}

\addplot [semithick, dashed, color2,line width=1.0pt, mark=square, mark size=2, mark options={solid}]
table {%
25.2562274932861 0.00565252304077148
26.2565135955811 0.00145251750946045
27.2523784637451 0.000342521715164185
28.2563056945801 9.00185585021973e-05
29.2245388031006 3.00228595733643e-05
30.1936225891113 0.000170022249221802
31.1075382232666 0.00179502964019775
32.1059085845947 0.0133425176143646
};
\addlegendentry{\tiny ILA-NN, $F_s^{\text{ADC}}=220$ MHz}

\addplot [semithick, color2, mark=square, line width=1.0pt, mark size=2, mark options={solid}]
table {%
25.2598934173584 0.00531253218650818
26.266357421875 0.0013225257396698
27.2605285644531 0.000282526016235352
28.248067855835 6.00270080566406e-05
29.2330265045166 1.67632102966309e-05
30.1886806488037 0.000145035982131958
31.1109504699707 0.00169003009796143
32.1143676757812 0.0121125221252441
};
\addlegendentry{\tiny ILA-NN full-rate~\cite{wu2020residual}}

\addplot [semithick, color1, mark=triangle, line width=1.0pt,mark size=2, mark options={solid}]
table {%
25.296573638916 0.00465752482414246
26.2886199951172 0.00142002701759338
27.2849636077881 0.000287520885467529
28.2901649475098 7.00261783599853e-05
29.2506684875488 2.10084495544434e-05
30.1735137939453 0.000180030546188354
31.1451686859131 0.00200252566337585
32.1354179382324 0.0177550327777863
};
\addlegendentry{\tiny ILA-GMP full-rate~\cite{GMP_2006}}

\addplot [semithick, color3,line width=1.0pt, mark=o, mark size=2, mark options={solid}]
table {%
25.26975440979 0.00511877834796906
26.2696094512939 0.00148402988910675
27.2701358795166 0.000301026525497437
28.2618541717529 5.55263662338257e-05
29.2455177307129 1.55143947601318e-05
30.2441940307617 0.000119776887893677
31.2243843078613 0.0010592782497406
32.2240180969238 0.0055727744102478
};
\addlegendentry{\tiny RL-NN}

\addplot [semithick, color4, line width=1.0pt, mark=star, mark size=2.5, mark options={solid}]
table {%
25.2974853515625 0.0047100305557251
26.2941722869873 0.00128003358840942
27.2932472229004 0.000277525186538696
28.2946510314941 4.00185585021973e-05
29.2912330627441 4.50339508056641e-06
30.2974948883057 3.25186347961426e-05
31.2706146240234 0.000361774563789368
32.2142143249512 0.00277652621269226
};
\addlegendentry{\tiny Ideal PA~\cite{chani2018lower}}

\addplot [semithick, black, line width=1.0pt, mark options={solid,rotate=180}]
table {%
25.2972316741943 0.00482953257359764
26.2893009185791 0.00138773834785744
27.2864665985107 0.000293452161519214
28.2929439544678 4.16958631942466e-05
29.296802520752 3.67520504640506e-06
30.2867870330811 1.8370367471654e-07
31.2731609344482 4.18556322934194e-09
32.2048988342285 3.94466681541417e-11
};
\addlegendentry{\tiny $64$ QAM Theory}
\end{axis}

\end{tikzpicture}
	\vspace*{-0.1 \baselineskip}
	\caption{SER as a function of the average PA output power.}
	\label{fig:ser_Pavg_gmp}
	\vspace*{-0.5 \baselineskip}
\end{figure}

Fig.~\ref{fig:ser_Pavg_gmp} shows the testing \ac{ser} results versus the average PA output power for the cases of no DPD, ILA-optimized NN DPD~\cite{wu2020residual} with under-sampling ($F_{s}^{\text{ADC}}=110$ and $220$ MHz) and full-rate ($F_{s}^{\text{ADC}}=550$ MHz) ADCs in the feedback path, ILA-optimized GMP DPD with full-rate  ADC~\cite{GMP_2006}, the proposed RL-optimized DPD, the case with linear-clipping PA~\cite{chani2018lower},\footnote{The linear-clipping PA has a linear behavior before the clipping region, which has the minimum distortion that any DPDs can achieve.} and the theoretical SER bound of $64$ \ac{qam}, where the average energy per bit is converted to the average PA output power considering a distortion-free PA. The RL-optimized \ac{dpd} is optimized using the  cross-entropy loss function.  Note that the policy $\pi(\cdot)$ is removed once the training ends, and the input of the PA becomes $\boldsymbol{x}$. The number of QAM symbols used to calculate the SER is $10^7$.

We observe that the \ac{ser} curves exhibit two different behaviors. When the average PA output power is below $29$ dBm, most of the PA input signal $\boldsymbol{x}$ is in the linear region of the PA, and the SER of all scenarios with DPD improve quickly with increasing  average PA output power $P_{\text{out}}$. The proposed RL-optimized NN DPD achieves similar SER as the ILA-optimized NN and GMP DPDs with full-rate ADCs, and has substantial SER gain over the ILA-optimized NN DPDs with undersampling ADCs. As the average PA output power increases above $29$ dBm, $\boldsymbol{x}$ starts to be clipped by the saturation region of the PA. While the clipping effect makes the SER of all DPD cases degrade rapidly, we note that the RL-optimized DPD exhibits considerable SER gains over other ILA-optimized DPDs at the highly nonlinear region of the PA, i.e., $30$ dBm $<P_{\text{out}}<32$ dBm. This indicates the advantage of the symbol-based criterion for DPD optimization over the sample-based criterion in terms of SER.  Overall, the results prove the effectiveness of the sample-based DPD optimization using a symbol-based criterion without knowing the hardware and channel for a PA with memory and the \ac{awgn} channel. The SER gap between all the DPD cases and the linear-clipping case may come from some residual distortions due to irreversible nonlinearity.

\subsubsection{In-band and Out-of-band Errors}
\begin{figure}[t]
    \centering
    \input{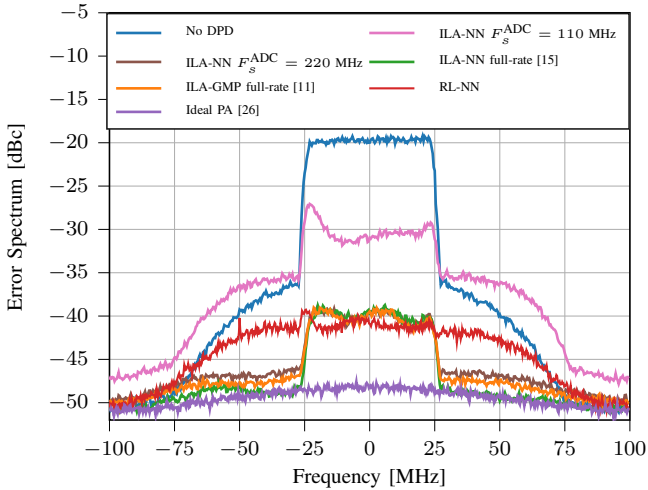}
    \caption{Error spectrum between the actual and desired PA output signals of schemes in Fig.~\ref{fig:ser_Pavg_gmp} at the average PA output power $P_{\text{out}}=30.2$ dBm.}
    \label{fig:psd}
\end{figure}

Fig.~\ref{fig:psd} shows the error spectrum (i.e., the \ac{psd} of the difference between the real and ideal PA output signals) of schemes in Fig.~\ref{fig:ser_Pavg_gmp} at the average PA output power $P_{\text{out}}=30.2$ dBm.  The corresponding \ac{nmse} and \ac{acpr} results are presented in Table~\ref{tab:psd}.

Due to the aliasing and band-limiting effects of the undersampling ADC, the linearization performance of the ILA-NN DPD with $F_{s}^{\text{ADC}}=110$ MHz is affected severely, exhibiting large in-band and out-of-band errors compared with DPDs of full-rate ADCs and even larger out-of-band errors ($-35.4$ dBc ACPR) compared with No DPD case ($-37$ dBc ACPR). With a full-rate ADC, the power spectral errors of the ILA-NN are reduced ($-38.7$ dB NMSE and $-40.2$ dBc ACPR). As expected,  the in-band spectral error of the RL-NN-based DPD is better than its out-of-band spectral error since the optimization criterion, i.e., symbol-based criterion, focuses more on the in-band errors. Nevertheless, the RL-NN-based DPD still maintains a satisfactory ACPR ($-38.1$ dBc) considering that the lower bound is $-41.3$ dBc.
\begin{table}[t]
\centering
\caption{NMSE and ACPR results of cases in Fig.~\ref{fig:psd} at the average PA output power $P_{\text{out}}=30.2$ dBm.}
\begin{tabular}{@{}ccc@{}}
\toprule
 & NMSE [dB] & ACPR [dBc]  \\
 \midrule 
 No DPD &$-22.8$ & $-37.0$ \\
 ILA-NN, $F_s^{\text{ADC}}=110$ MHz &$-28.3$ &$-35.4$\\
  ILA-NN, $F_s^{\text{ADC}}=220$ MHz &$-37.9$ &$-39.2$\\
 ILA-NN, full-rate~\cite{wu2020residual} &$-38.7$ &$-40.2$\\
 ILA-GMP, full-rate~\cite{GMP_2006} & $-38.2$&$-39.8$\\
 RL-NN  & $-35.0$ & $-38.1$\\
 Ideal PA~\cite{chani2018lower} &$-42.3$ & $-41.3$
 \\    
\bottomrule
\end{tabular}
\label{tab:psd}
\end{table}

\section{Conclusion}
We proposed a novel DPD optimization algorithm that optimizes DPD parameters using a symbol-based criterion at the receiver side instead of a sample-based criterion at the transmitter side, which avoids the cumbersome feedback path in the transmitter. The proposed optimization algorithm, based on RL, does not require a model for the hardware or channel, which is an attractive feature in practice. Exploiting the policy gradient theorem, we connect the symbol-based criterion with the sample-based policy optimization. The proposed algorithm is verified by simulation results for a \ac{gmp} PA modeled from a real PA over the AWGN channel. The proposed RL-NN-based DPD achieves SER gains over the ILA-based DPD even the latter uses a full-rate ADC in the feedback path, while it also maintains satisfactory out-of-band errors. It is expected to see a performance improvement for a real PA as the performance of ILA is limited by the nature of its identification process. Some limitations of this work are the AWGN channel and hardware models, which can be generalized to a more realistic scenario considering more realistic channels and more hardware impairments (e.g., quadrature imbalance).

\balance

\bibliographystyle{IEEEtran}

\bibliography{./bibliography/IEEEabrv,reference_list}
\end{document}